\documentclass[twocolumn,10pt,nofootinbib,superscriptaddress]{revtex4-2}
\usepackage{graphicx} 
\usepackage{amsmath}
\usepackage{color}
\usepackage{orcidlink}
\usepackage{siunitx} 
\usepackage{footmisc}
\usepackage{amssymb}
\usepackage{placeins}
\usepackage{hyperref}

\DeclareSIUnit{\gauss}{G} 

\begin{document}
\title{NASDUCK$^\prime$: Laboratory Limits on Ultralight Dark-Photon Dark Matter with Null-Axis Magnetometry} 
\author{Joel Barir}
\affiliation{School of Physics and Astronomy, Tel Aviv University, Tel Aviv, Israel}
\author{Itay M. Bloch \orcidlink{0000-0003-1931-4344} }
\affiliation{Theoretical Physics Department, CERN, 1 Esplanade des Particules, CH-1211 Geneva 23, Switzerland}
\affiliation{Physics Department, Technion – Israel Institute of Technology, Haifa 3200003, Israel}
\author{Yair Goldszeft}
\affiliation{School of Physics and Astronomy, Tel Aviv University, Tel Aviv, Israel}
\author{Gal Goldstein}
\affiliation{Rafael Ltd., 31021 Haifa, Israel}
\author{Constantine Feinberg \orcidlink{0000-0001-8761-6585}}
\affiliation{School of Physics and Astronomy, Tel Aviv University, Tel Aviv, Israel}
\author{Or Katz\orcidlink{0000-0001-7634-1993}}
\affiliation{School of Applied and Engineering Physics, Cornell University, Ithaca, NY 14853}
\author{Tomer Volansky}
\affiliation{School of Physics and Astronomy, Tel Aviv University, Tel Aviv, Israel}

\begin{abstract}
The dark photon is a well-motivated ultralight dark-matter candidate that may couple to the Standard Model through kinetic mixing. We search for dark-photon dark matter in the mass range $m_{A'}c^2 = 4\times10^{-12}$--$2\times10^{-9}\,\mathrm{eV}$ (1--500~kHz) using a three-axis magnetometer inside a large conductive shielded room. We set new laboratory limits on the kinetic-mixing parameter $\epsilon$, improving upon previous laboratory bounds by up to three orders of magnitude. Our search exploits a geometry-defined null response along one axis as a noise reference; a subtraction procedure reduces the noise floor and improves sensitivity. These results establish the strongest laboratory constraints in this mass range and illustrate how null-axis magnetometry can broaden terrestrial searches for ultralight vector dark matter.
\end{abstract}
 \maketitle

\section{Introduction}
Astrophysical and cosmological observations provide compelling evidence for dark matter (DM), yet its microscopic identity remains unknown. Ultralight bosonic DM with masses $m \lesssim \SI{1}{\electronvolt\per c\squared}$ is well described as a coherently oscillating classical field, since the observed local DM energy density implies a large occupation number~\cite{Arias:2012az,Nelson:2011sf,Ferreira:2020fam,Hui:2016ltb}. Such fields can produce narrowband, time-varying laboratory signatures even for extremely weak couplings to the Standard Model. Axion-like particles and dark photons are canonical examples~\cite{Jaeckel:2010ni,Fabbrichesi:2020wbt,PDG2024:Axions}.

Dark photons arise as a vector boson that kinetically mixes with the Standard Model photon~\cite{Fabbrichesi:2020wbt,HOLDOM1986196}. Kinetic mixing makes a dark-photon DM field appear as an effective oscillatory current density that sources ordinary electromagnetic fields at a frequency set by the dark-photon mass~\cite{Chaudhuri:2014dla}. Because this source term is present throughout space, it permeates conductive shielding even as external electromagnetic radiation is excluded.

Conductive shielding plays a dual role: in addition to suppressing environmental backgrounds, it fixes the electromagnetic boundary conditions that determine the in situ response to the effective source. The resulting magnetic-field signal is set by the shield geometry and, in the long-wavelength regime, scales with the characteristic size of the enclosure~\cite{Chaudhuri:2014dla}. This motivates searches in large, well-characterized shielded enclosures, where the response is calculable and the attainable dark-photon signal is enhanced.

A broad range of methods have been proposed to search for ultralight DM~\cite{ADMX:2010ubl,Jaeckel:2007ch,Redondo:2010dp,Bahre:2013ywa,Baryakhtar:2018doz,Gelmini:2020kcu,Horns:2012jf,Suzuki:2015sza,FUNKExperiment:2020ofv,Ramanathan:2022egk,An:2022hhb}.  A representative example for shielding-based searches is the DM-Radio concept, which uses a tunable $LC$ resonator inside a conductive shield to sense the induced magnetic response in a controlled geometry~\cite{Chaudhuri:2014dla,Phipps:2019cqy}. Implementations span from geophysical-scale shielding at $\lesssim\SI{10}{\hertz}$~\cite{Fedderke:2021aqo,Fedderke:2021rrm,Sulai:2023zqw,Nomura:2025rfi} to shielded magnetometer searches at $\SIrange{1}{500}{\hertz}$~\cite{Jiang:2023jhl} and higher-frequency electric-field measurements~\cite{Godfrey:2021tvs}. The intermediate $\SIrange{1}{500}{\kilo\hertz}$ band remains comparatively less explored by laboratory measurements: large shielded enclosures enhance the predicted response, but residual magnetic backgrounds can limit sensitivity.

Here we search for dark-photon DM in the $\SIrange{1}{500}{\kilo\hertz}$ band using a three-axis magnetometer in a large electromagnetically shielded room and set new laboratory limits on the kinetic-mixing parameter $\epsilon$. We use the shield-defined signal geometry as a built-in noise reference: at the sensor location, the predicted dark-photon response is near null along one measurement axis, which directly monitors residual noise. Using this null-axis channel, we subtract shared low-frequency noise and extend sensitivity in the noise-dominated regime while maintaining response to a dark-photon signal. The resulting constraints are the strongest laboratory limits in this mass range and surpass bounds from Coulomb-law tests by two to three orders of magnitude~\cite{Bartlett:1988yy}.

\begin{figure*}[t]
\centering
\includegraphics[width=\linewidth]{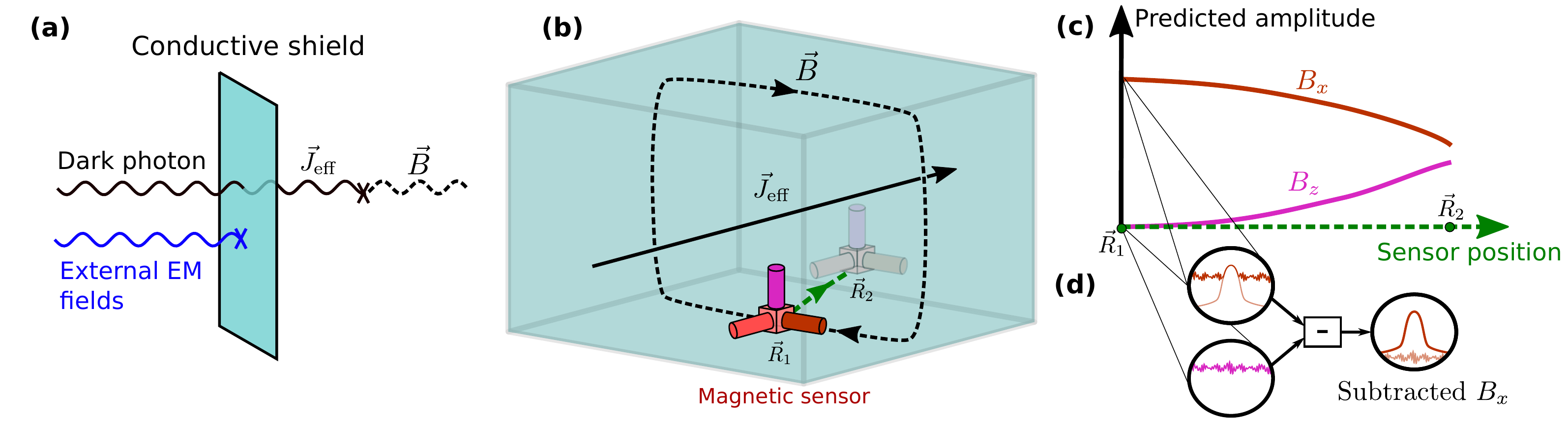}
\caption{Detection concept and null-axis subtraction. \textbf{(a)} Conductive shielding suppresses external electromagnetic fields, while a dark-photon DM field
effectively persists inside the shielded volume. The effect of the dark photon vector field
$\vec{A}$ is equivalent to an effective current $\vec{J}_\mathrm{eff}= -\epsilon m_{A'}^2 \vec{A'}$, which generates a magnetic field inside the shield. \textbf{(b)} $\vec{J}_\mathrm{eff}$ generates a magnetic field $\vec{B}$ that is strongest near the faces parallel to it. A magnetic sensor is placed at the center of the floor (point $\vec{R}_1$), where magnetic fields will be excited by the $x$ and $y$ components of $\vec{J}_\mathrm{eff}$. \textbf{(c)} As the detector is moved from point $\vec{R}_1$ to point $\vec{R}_2$, the signal strength in the $x$ and $z$ directions changes.
\textbf{(d)} We measure $\vec{B}$ at point $\vec{R}_1$, where the signal in the $z$ direction vanishes identically, regardless of the amplitude and direction of $\vec{J}_\mathrm{eff}$. In the $x$ direction, the dark photon produces a narrow-band signal in Fourier space, with frequency determined by the dark photon's rest mass. The $z$-direction measurement, serving as a pure noise channel, is subtracted
from the measurement in the $x$ direction, yielding a lower noise version of it without affecting the signal. The same subtraction procedure is performed for $B_y$, not shown here.}
\label{fig:drawing}
\end{figure*}

\section{Dark Photon Interaction with Matter}
\label{sec:photon_current}
Kinetic mixing between the dark photon $A'_\mu$ and the Standard Model photon $A_\mu$ makes a dark-photon DM field act as an effective classical source for ordinary electromagnetism~\cite{Chaudhuri:2014dla,Graham:2014sha,Fedderke:2021aqo,Fabbrichesi:2020wbt}. In the interaction basis, this is captured by an effective four-current $J_{\mathrm{eff}}^{\mu}\equiv -\epsilon m_{A'}^{2}A'^{\mu}$; for nonrelativistic galactic DM ($|A'^0|\ll|\vec A'|$) the relevant source is $\vec J_{\mathrm{eff}}=-\epsilon m_{A'}^{2}\vec A'$. The field oscillates at $\omega_{A'}\simeq m_{A'}c^{2}/\hbar$ and is narrowly distributed by the halo velocity dispersion, with fractional bandwidth $\Delta \omega_{A'}/\omega_{A'}\sim10^{-6}$ and amplitude $|\vec A'|\sim\hbar \sqrt{2 \rho_{\rm DM}}/m_{A'}$ in the Standard Halo Model, where $\rho_{\mathrm{DM}}$ is the DM density (App.~\ref{app:DM_model}). The polarization of $\vec A'$ (and hence the direction of $\vec J_{\mathrm{eff}}$) is unknown; we assume individual DM particles have independently oriented polarizations. A violation of this assumption would modify the inferred bound, but we expect the effect to be modest, as our experiment is simultaneously sensitive to two orthogonal components of $\vec{J}_{\mathrm{eff}}$ and located at a latitude that is nearly optimal for detecting a fixed-polarization dark photon signal~\cite{Caputo:2021eaa}.

Inside a conductive shield, external electromagnetic fields are strongly attenuated, while the effective source persists and drives fields within the enclosure~\cite{Chaudhuri:2014dla}. In our search band $\omega_{A'}\ll c/L$, so $\vec J_{\mathrm{eff}}$ may be treated as spatially uniform across the interior. Solving Maxwell's equations with conducting boundary conditions then yields a geometry-dependent magnetic response transverse to $\vec J_{\mathrm{eff}}$. For a rectangular room and a sensor placed at the center of a face (Fig.~\ref{fig:drawing}), the two in-plane components of $\vec J_{\mathrm{eff}}$ map onto orthogonal magnetic-field components at the sensor, with characteristic signal amplitude
\begin{equation}
\label{eq:B_scaling}
B_{\mathrm{sig}}\simeq \epsilon\,\mathcal{G} \frac{c^2}{\hbar}\,\omega_{A'}L\,\sqrt{\rho_{\mathrm{DM}}},
\end{equation}
where $L$ is a characteristic room dimension and $\mathcal{G}$ is a dimensionless geometry factor set by the enclosure and sensor location; for our configuration $|\mathcal{G}|\approx 0.5$ averaged over all current directions (see App.~\ref{app:B_full}). Equation~\eqref{eq:B_scaling} highlights the key scaling in the ultralight regime: the predicted magnetic response grows linearly with the enclosure size.

Crucially, the same boundary conditions enforce a near-null dark-photon response in the vertical component at the sensor location, $B_z\simeq 0$ in the idealized geometry for any direction of $\vec J_{\mathrm{eff}}$ (App.~\ref{app:B_full} and Fig.~\ref{fig:drawing}). Measuring $B_z$ therefore provides a dedicated noise monitor, while $B_x$ and $B_y$ serve as signal channels. This null-axis feature underpins the subtraction procedure introduced in Sec.~\ref{sec:analysis}, enabling improved sensitivity in correlated noise-dominated frequency ranges.

\section{Experimental Setup}
We measure the time-varying magnetic field with a three-axis search-coil magnetometer (LEMI-150)~\cite{LEMI150_ISR}. The three outputs are digitized with a National Instruments USB-6341-BNC at $\SI{1}{\mega\hertz}$, covering the $\SIrange{1}{500}{\kilo\hertz}$ search band. We use low-pass filters to mitigate aliasing and high-pass filters with a cuttoff at $\SI{1}{\kilo\hertz}$. The typical noise level of this setup is $10\,\mathrm{fT}/\sqrt{\mathrm{Hz}}$ at frequencies above $\SI{100}{\kilo\hertz}$.

We determine the end-to-end, frequency-dependent transfer function of each axis by applying a known magnetic field and measuring the sensor output, thereby converting the digitized voltage spectra into calibrated magnetic-field spectra across the full band. From $\SI{1}{\kilo\hertz}$ to $\SI{50}{\kilo\hertz}$, the reference field is generated with an industrial Helmholtz coil (Schwarzbeck Mess-Elektronik, model 5216). From $\SI{50}{\kilo\hertz}$ to $\SI{500}{\kilo\hertz}$, we use a custom-wound square coil and normalize its response to the Helmholtz-coil calibration. The resulting transfer functions are used throughout the analysis.

Measurements are performed inside a conductive shielded room of dimensions $10\,\mathrm{m}\times6\,\mathrm{m}\times8\,\mathrm{m}$, with an inner $\SI{2}{\milli\meter}$ aluminum layer and an outer $\SI{3}{\milli\meter}$ galvanized-steel layer separated by a $\SI{40}{\milli\meter}$ air gap. We verified shielding by comparing magnetic-field spectra recorded outside and inside the room in the $\SIrange{80}{180}{\kilo\hertz}$ band, using narrowband ambient transmissions as probes, and found an attenuation of the order of $10^{-5}$ in magnetic-field amplitude (App. \ref{app:Room}). The magnetometer is placed near the center of the floor to maximize sensitivity to the in-plane dark-photon response (Sec.~\ref{sec:photon_current} and App.~\ref{app:B_full}) and elevated by $\SI{20}{\centi\meter}$. The system is powered by lead-acid batteries to suppress line-frequency pickup.

\section{Data acquisition and analysis} \label{sec:analysis}
We recorded a two-hour continuous time-domain dataset from three orthogonal axes. To prevent analysis bias, we fixed all analysis choices before unblinding this dataset by validating the analysis chain on dedicated simulations and on an independent 10~min data subset excluded from the final result. Simulated data include white noise and injected dark-photon signals drawn from the Standard Halo Model (SHM), demonstrating robust signal recovery and validating the limit-setting procedure (App.~\ref{app:analysis}).

For each channel we Fourier transform the digitized voltages and apply the measured, frequency-dependent transfer function to obtain calibrated magnetic-field spectra. We then search for a stochastic SHM signal, which is distinguishable from instrumental backgrounds by its narrow fractional linewidth $\Delta f/f\sim10^{-6}$. We scan dark-photon masses and, at each mass, fit a signal-plus-noise model using a maximum-likelihood statistic (App.~\ref{app:analysis}). We first obtain ``direct" limits using only the in-plane channels $B_x$ and $B_y$.

To improve sensitivity in correlated noise bands, we exploit the null-axis property: at our sensor location the expected dark-photon response in $B_z$ is strongly suppressed, so $B_z$ provides a coherent monitor of residual instrumental noise. We construct a frequency-dependent complex transfer function $H_{z\to x}(f)$ (and similarly $H_{z\to y}(f)$) from window-averaged cross spectra between $B_z$ and the signal channel,
\begin{equation}
    H_{z\to x}(f) \equiv \frac{\langle B_{x}(f) {B_{z}}^{*}(f)\rangle}{\langle |B_{z}(f)|^2 \rangle},
\end{equation}
where $\langle\cdots\rangle$ denotes an average over a frequency window centered at $f$ and with a width of  10 kHz, which is much wider than the SHM linewidth.  Our
results are insensitive to reasonable variations of this width. We then form noise-subtracted spectra by removing the correlated component,
\begin{equation}
\label{eq:subtraction}
B_{x}^{(\mathrm{sub})}(f)=B_x(f)-H_{z\to x}(f)\,B_z(f),
\end{equation}
and analogously for $B_y$. To accommodate slow drifts, $H(f)$ is estimated and applied independently in 10~min segments, which are then recombined. The noise-subtracted data are processed through the same likelihood analysis, yielding the final constraints.

A potential concern is that a small dark-photon signal present in $B_z$, due to imperfect sensor positioning or instrumental crosstalk, could be subtracted from $B_x$. We bound this effect quantitatively.
First, the idealized room response suppresses the dark-photon amplitude in $B_z$ by a factor $\gtrsim 20$ relative to the signal channels at our sensor position (App.~\ref{app:B_full}). Second, the measured subtraction gain satisfies $|H_{z\to x}(f)|<2.5$ (and similarly for $H_{z\to y}$) across the full band. The maximum fractional signal removed by the subtraction term in Eq.~\eqref{eq:subtraction} is therefore bounded by
$|H_{z\to x}|/20 \lesssim 0.13$,
i.e., at most $\sim 10\%$ of the true signal amplitude under conservative assumptions.

We further validate the subtraction and limit-setting procedure with end-to-end signal injections passed through the full analysis chain, confirming recovery of SHM-like signals and bounding any residual attenuation under conservative injected leakage into $B_z$ (App.~\ref{app:analysis}).

The likelihood model assumes white noise plus a possible SHM signal; in practice, narrow coherent spectral lines violate this assumption and can generate spurious candidates. Candidate masses are first identified in the direct (unsubtracted) scan. We then apply three vetoes defined before unblinding: (1) a lineshape consistency test against the SHM template, rejecting features narrower than the expected DM linewidth~\cite{Bloch:2022kjm}; (2) rejection of candidates with coincident peaks in the nominally signal-suppressed $B_z$ channel; and (3) rejection of candidates excluded by the stronger limits obtained from the noise-subtracted data.
Vetoes (2) and (3) both rely on the null-axis $B_z$ measurement but are complementary in scope. Veto (2) targets narrowband coincidences in $B_z$, rejecting candidates even when the underlying disturbance is narrower than the predicted dark-photon signal. By contrast, veto (3) exploits the wideband correlations removed by subtraction, which suppress common external noise sources broader than the searched signal (details in App.~\ref{app:analysis}).

\begin{figure*}[t]
    \centering
    \includegraphics[width=16cm]{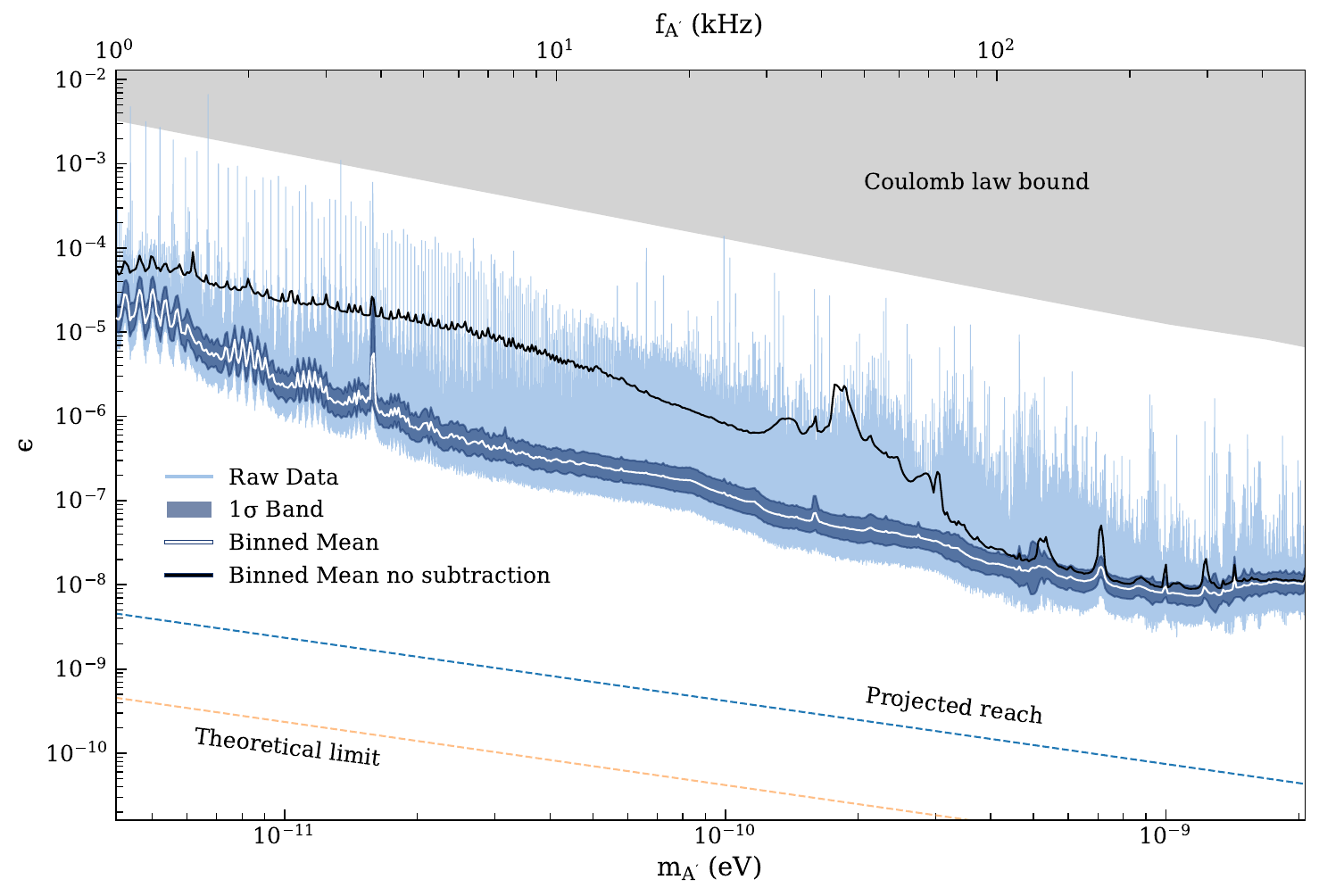}
    \caption{95\% C.L. upper limits on the kinetic-mixing parameter $\epsilon$ versus dark-photon mass $m_{A'}$ from this work. Limits are derived from the $B_x$ and $B_y$ channels, with sensitivity enhanced by null-axis subtraction using $B_z$ as a coherent noise reference (Fig.~\ref{fig:drawing}). The {\bf light-blue} region shows the point-by-point noise-subtracted limit ($>10^7$ scanned masses); the {\bf white} curve is a 1\% logarithmic-space average and the {\bf dark-blue} band indicates the corresponding standard deviation. The {\bf black} curve shows the averaged ``direct'' limit obtained without subtraction. Also shown are the Coulomb-law bound ({\bf gray})~\cite{Bartlett:1988yy}, an illustrative projection for a $(50\,\mathrm{m})^3$ implementation with $0.1\,\mathrm{fT}/\sqrt{\mathrm{Hz}}$ sensitivity ({\bf light-blue dashed})~\cite{Gramolin:2020ict}, and the corresponding Johnson-noise floor ($\sim10^{-2}\,\mathrm{fT}/\sqrt{\mathrm{Hz}}$, {\bf orange dashed}). Astrophysical limits can be stronger at comparable masses~\cite{Caputo:2021eaa} but are subject to modeling uncertainties; see Ref.~\cite{Hook:2025pbn}.}
    \label{fig:results}
\end{figure*}

\section{results}
Our primary result is a 95\% confidence-level upper limit on the kinetic-mixing parameter $\epsilon$ as a function of dark-photon mass $m_{A'}$, shown in Fig.~\ref{fig:results} for $m_{A'}c^2 = 4\times10^{-12}$--$2\times10^{-9}\,\mathrm{eV}$ ($\SIrange{1}{500}{\kilo\hertz}$). Limits are obtained with a likelihood-ratio construction at each scanned mass (App.~\ref{app:analysis}). The light-blue curve shows the limits after null-axis subtraction; the white curve is their local logarithmic average, with the dark-blue band indicating the corresponding standard deviation. The black curve shows the binned average limit from the direct (non-subtracted) analysis. Subtraction provides a major improvement in regions dominated by channel-correlated noise, and agrees with the direct analysis where such correlations are negligible.

Narrow coherent spectral features generated several tens of thousands of local outliers in the likelihood scan, most of which were subsequently rejected by the vetoes defined in Sec.~\ref{sec:analysis}. One remaining mass near the low-frequency edge of the search band survived all vetoes and is discussed in App.~\ref{app:analysis}. This feature appeared near the low-frequency edge of the search band, where the spectral-shape test is intrinsically less discriminating for a measurement duration of 2~hours, as only a small number of Fourier bins lie within the expected signal bandwidth.
These features do not affect the validity of the reported upper limits.

Fig.~\ref{fig:results} compares our laboratory constraints with the most stringent previous terrestrial bound in this mass range from precision tests of Coulomb's law (gray)~\cite{Bartlett:1988yy}. Our limit strengthens existing laboratory constraints by up to three orders of magnitude, establishing the strongest laboratory bounds in this mass range. Astrophysical constraints can be stronger at comparable masses~\cite{Caputo:2021eaa}, but are subject to modeling assumptions and systematic uncertainties; recent work has highlighted potential limitations of the dominant astrophysical bounds~\cite{Hook:2025pbn}. Our result therefore provides a robust and complementary terrestrial constraint.

Also shown in Fig.~\ref{fig:results} is an illustrative projected reach for a scaled-up implementation (light-blue dashed), assuming a shielded volume of $(50\,\mathrm{m})^3$ and a magnetic-field sensitivity of $0.1\,\mathrm{fT}/\sqrt{\mathrm{Hz}}$, consistent with state-of-the-art axion searches~\cite{Gramolin:2020ict}. The orange dashed curve indicates the corresponding Johnson-noise floor arising from the surrounding conductive environment; for a shield resistivity of $3\,\Omega\cdot \mathrm{m}$ this corresponds to a magnetic noise level of $\sim10^{-2}\,\mathrm{fT}/\sqrt{\mathrm{Hz}}$. Higher resistivity reduces Johnson noise but simultaneously weakens electromagnetic shielding, allowing external low-frequency fields to penetrate and ultimately limiting achievable sensitivity.

\section{Summary and Prospects}
We have performed a laboratory search for ultralight dark-photon dark matter using three-axis magnetometry inside a large conductive shielded room, and set new 95\% C.L. upper limits on the kinetic-mixing parameter $\epsilon$ over $m_{A'}c^2 = 4\times10^{-12}$--$2\times10^{-9}\,\mathrm{eV}$ ($\SIrange{1}{500}{\kilo\hertz}$). A key advance is a null-axis subtraction technique that uses the strongly signal-suppressed channel as a coherent noise reference, enabling improved sensitivity in portions of the spectrum dominated by correlated noise. These results establish the strongest laboratory constraints in this mass range.

The null-axis approach is broadly applicable to shielding-based searches in which conducting boundary conditions enforce a suppression of one field component at the sensor location. Looking forward, the accessible mass range can be shifted or widened with sensors optimized for different frequency bands, while sensitivity for ultralight masses can be improved by increasing the shielded volume, and by reducing instrumental and environmental magnetic noise toward the Johnson-noise floor. A related direction is to implement null-axis referencing in resonant architectures (e.g., cavity or $LC$ configurations), where the dark-photon-induced response can be enhanced within a well-controlled mode structure. More generally, analogous null-channel referencing may be adaptable to other ultralight dark-matter searches that rely on correlated electromagnetic readout, including axion-like particle experiments. Together, these avenues outline a scalable terrestrial program for ultralight dark matter that is both systematically controlled and complementary to astrophysical probes.

{\bf Acknowledgments.}  OK and TV acknowledge support from the Binational Science Foundation (grant No.\ 2024160). TV is supported, in part, by the Israel Science Foundation (grant No.\ 2803/25).
CF and TV are jointly supported by the PAZY Foundation (grant No.\ 429).   JB\ thanks the Alexander Zaks scholarship for its support.

\clearpage
\onecolumngrid
\appendix

\section{Statistical model of dark matter} \label{app:DM_model}

We present here the stochastic model of the dark photon field used in the data analysis. The construction closely follows that of a previous NASDUCK work~\cite{Bloch:2021vnn}, but is adjusted to model a vector dark photon field instead of axions.

First, we write the contribution of a single dark photon particle to the field, in the rest frame of the sun:
\begin{equation}
	\vec{A}'(t)|_{\rm sp} =  \frac{\vec{e}}{m_{\rm DM}}
	\sqrt{\frac{\rho_{\rm DM}}{2 N_{\rm DM}}} e^{iE_{\rm DM}t /\hbar + i\phi_{\rm DM}}
	+ \text{h.c.}
\end{equation} 
where $E_{\rm DM} = m_{\rm DM}c^2 + m_{\rm DM}v^2 /2$ is the particle's energy, $\vec{v}$ its velocity, and $\vec{e}$ is a unit vector describing its polarization. The amplitude is normalized by the total DM density $\rho_{\rm DM}$ and the number of particles $N_{\rm DM}$ such that the total energy density of the field equals $\rho_{\rm DM}$. 
The plane-wave factor $\vec{k}_{\rm DM}\cdot\vec{x}$ in the exponent has been replaced with a constant random phase $\phi_{\rm DM}$, since within the rest frame of the sun, it does not meaningfully change during data-taking.

The variables $\vec{e}$, $v$, and $\phi_{\rm DM}$ are treated as random variables, independently distributed for each particle. 
A violation of the assumption of independent polarizations $\vec{e}$ would modify the inferred bound, but we expect the effect to be modest, as our experiment is simultaneously sensitive to two orthogonal components of $\vec{J}_{\mathrm{eff}}$ and located at a latitude that is nearly optimal for detecting a fixed-polarization dark photon signal~\cite{Caputo:2021eaa}.
The polarization $\vec{e}$ is uniformly distributed on the unit sphere, the phase $\phi_{\rm DM}$ is uniformly distributed on the unit circle, and the velocity $\vec{v}$ is distributed according to the Standard Halo Model~\cite{Lee:2013xxa}. The model provides the probability distribution function for vector $\vec{v}$:
\begin{equation} \label{eq:SHM}
	f_{\mathrm{SHM}}(\vec{v})=
	\begin{cases}\frac{1}{N_{\mathrm{SHM}}} e^{-(\vec{v}-\vec{v}_{\odot})^2/v_{\mathrm{vir}}^2}
	& |\vec{v}-\vec{v}_{\odot}|<v_{\mathrm{esc}} \\ 0 &,
\text { otherwise,}\end{cases}  
\end{equation}
where $v_{\rm vir} = \SI{220}{\kilo\meter\per\second}$ is the virial velocity, $v_{\rm esc}=\SI{540}{\kilo\meter\per\second}$ is the escape velocity, $\vec{v}_{\odot}= (0,0,232\mathrm{~km} / \mathrm{s})$ is the sun's velocity (chosen to be in the $\hat{z}$ direction), and $N_{\rm SHM}$ is a normalization factor ensuring that the distribution integrates to unity.

The total field is the sum over all particles:
\begin{equation} 
	\vec{A}'(t) =  \frac{1}{m_{\rm DM}}
	\sqrt{\frac{\rho_{\rm DM}}{2 N_{\rm DM}}} e^{im_{\rm DM}c^2t/\hbar}
	\sum_{j=1}^{N_{\rm DM}} \vec{e}_j\,e^{im_{\rm DM}v^2_jt /(2\hbar) +
	i\phi_{{\rm DM}, j}}
	+ \text{h.c.},
\end{equation}
where the variables for each particle are now labeled by the index $j$ indicating particle number. We now Fourier transform $\vec{A}'(t)$ to the frequency domain. For a finite measurement time $T$, the frequencies are discrete, and the Fourier components $\vec{A}'_n$ are given by:
\begin{equation} \label{eq:A_fourier}
\vec{A}'_n  =\frac{1}{m_{\rm DM}}\left[  \vec{W}_n\left(m_{\mathrm{DM}}, \Phi_{\mathrm{DM}}\right)+\vec{W}_n\left(-m_{\mathrm{DM}},-
	\Phi_{\mathrm{DM}}\right) \right]  ,
\end{equation} 
where we have defined
\begin{equation} \label{eq:W_n}
	\vec{W}_n\left(m_{\mathrm{DM}}, \Phi_{\mathrm{DM}}\right)=\sqrt{\frac{
\rho_{\mathrm{DM}}}{2N_{\mathrm{DM}}}}
\sum_{j=1}^{N_{\mathrm{DM}}}(-1)^n e^{i \left[ \phi_{\mathrm{DM}, j}+E_{\mathrm{DM}, j} T /(2 \hbar) \right] }
\operatorname{sinc}\left(\frac{\left( E_{\mathrm{DM}, j}
- \hbar \omega_n\right) T}{2 \hbar}\right) \vec{e}_j 
\end{equation}
with $\omega_n = 2\pi n / T$ being the discrete Fourier frequencies. For a large number of particles ($N_{\rm DM}\gg1$), the central limit theorem allows the probability distribution of $\vec{A}'_n$ to be approximated as a multivariate Gaussian.

For convenience in the statistical analysis, we define the dimensionless quantity $\bar{\mathbf{A}}$:
\begin{equation} \label{eq:A_normalization}
	\bar{\mathbf{A}}=\frac{m_{\rm DM}\vec{A}}{\sqrt{2 \rho_{\rm DM}} }.
\end{equation} 
To specify the Gaussian probability distribution, we must calculate the covariances between elements of $\bar{\mathbf{A}}$. These covariances are obtained by integrating over the probability distributions of the random variables in Eqs.~\eqref{eq:A_fourier} and \eqref{eq:W_n}. The real and imaginary components of $\bar{\mathbf{A}}_n$ are independent and identically distributed:
\begin{equation} \begin{gathered} 
 \operatorname{Cov}\left(\operatorname{Re}\left(\bar{A}_{n, i}\right),
	\operatorname{Re}\left(\bar{A}_{l,
	j}\right)\right)=\operatorname{Cov}\left(\operatorname{Im}\left(\bar{A}_{n,
	i}\right), \operatorname{Im}\left(\bar{A}_{l, j}\right)\right), \\
	\operatorname{Cov}\left(\operatorname{Re}\left(\bar{A}_{n, i}\right),
	\operatorname{Im}\left(\bar{A}_{l, j}\right)\right) = 0.
\end{gathered} \end{equation}
Since $\vec{A}(t)$ is real, it is sufficient to calculate the covariance for positive frequencies ($n,l > 0$):
\begin{equation} \label{eq:cov_integral}
	\operatorname{Cov}\left(\operatorname{Re}\left(\bar{A}_{n, i}\right),
	\operatorname{Re}\left(\bar{A}_{l, j}\right)\right) = 
	\frac{1}{6} \delta_{ij}\int_{}^{} d^3\vec{v}f_{\rm SHM}(\vec{v}) 
		(-1)^{n-l}\operatorname{sinc}\left( \frac{(E_{\rm DM}(v)-\hbar\omega_n)T}{2\hbar} \right) 
		\operatorname{sinc}\left( \frac{(E_{\rm DM}(v)-\hbar\omega_l)T}{2\hbar} \right).
\end{equation}
This expression leads to a spectrum that is sharply peaked. The peak is attained when the sinc arguments vanish within the integration range, and the peak frequency corresponds to the typical DM particle energy $\omega_{\mathrm{peak}}\approx m_{\mathrm{DM}}(c^2 + v_{\rm SHM}^2/2)/\hbar$ (where $v_{\rm SHM}$ refers to velocities appearing in Eq.~\eqref{eq:SHM}). The  width is determined by the kinetic energy distribution, which is governed by the velocity dispersion of the Standard Halo Model, leading to a relative bandwidth  of order $v_{\rm SHM}^2 /c^2 \approx 10^{-6}$. 
\begin{figure}[htpb]
	\centering
	\includegraphics[width=0.8\textwidth]{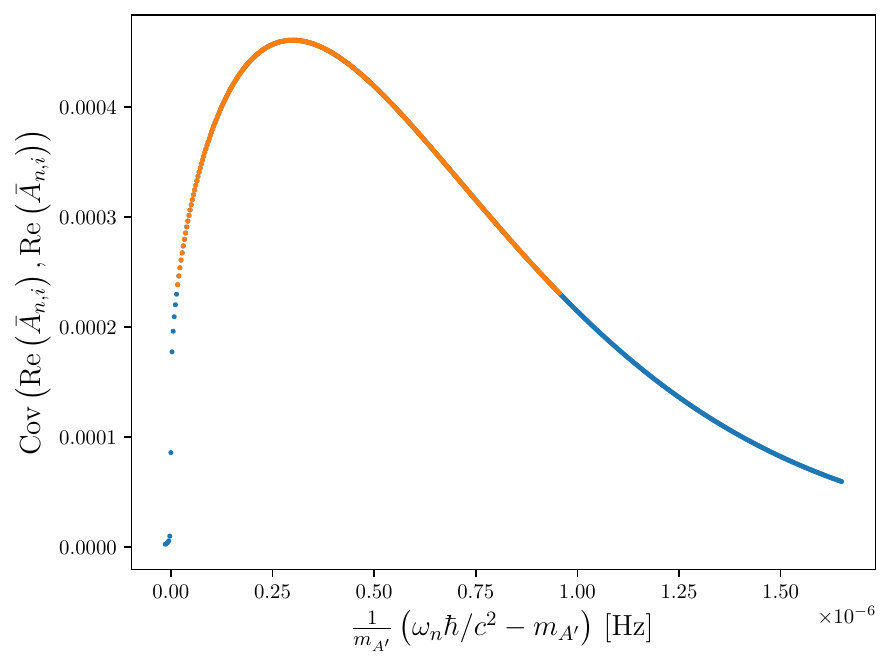}
	\caption{The diagonal elements of the dimensionless signal covariance matrix from Eq.~\eqref{eq:cov_integral}, illustrating the expected spectral shape of the dark photon signal. The parameters shown are for a signal at $f = m_{A'}c^2/h = \SI{50}{\kilo\hertz}$ with an integration time of $T=\SI{7200}{\second}$. \textbf{Orange points} indicate the frequencies above half of the maximum, included in the analysis for this specific mass hypothesis. 
    }
	\label{fig:cov}
\end{figure}
To illustrate the spectral shape, the diagonal elements of the covariance matrix are shown in Fig.~\ref{fig:cov}. These elements provide a good picture of the signal's spectral shape, as the off-diagonal elements, representing correlations between different frequencies, are significantly smaller. We define the analysis window as the frequency range where the  self covariance is above half of its maximum value. We have found with simulations that including in the likelihood analysis more frequencies beyond this range does not significantly improve
the result. As mentioned in the main text, masses were scanned with a relative resolution of $0.5\times 10^{-6}$--smaller than the analysis window, and tested by simulations to ensure sufficient overlap between neighboring masses.

The final Gaussian probability distribution for the dark photon field can be written as\footnote{This expression relies on the real and imaginary parts of $\bar{\mathbf{A}}$ being identically distributed an independent of each other.}:
\begin{equation} \label{eq:A_distro}
	P_{\bar{A}}(\bar{\mathbf{A}})=\frac{1}{(2 \pi)^{n_{\bar{A}}}
\left|\Sigma_{\bar{A}}\right|} e^{-\bar{\mathbf{A}}^{\dagger}
\Sigma_{\bar{A}}^{-1} \bar{\mathbf{A}} / 2} ,
\end{equation} 
where $n_{\bar{A}}$ is the total number of components in $\bar{\mathbf{A}}$, and $\Sigma_{\bar{A}}$ is its covariance matrix with size $n_{\bar{A}}\times n_{\bar{A}}$. The vector $\mathbf{\bar{A}}$ carries both a Fourier index $n$ and a spatial vector index $i$. Since our experiment is sensitive to two spatial components ($x$ and $y$), the likelihood analysis is restricted to those components, making the number of relevant elements $n_{\bar{A}} = 2n_f$, where $n_f$ is the number of  relevant Fourier frequencies included in the analysis (as defined above, and shown in Fig.~\ref{fig:cov}).

\section{Fields in Rectangular Shield} \label{app:B_full}
We present here the full form of the magnetic field $\vec{B}$ generated inside the rectangular room by the dark photon field $\vec{A'}$. As described in the main text, the dark photon is equivalent to an effective current $\vec{J}_{\rm eff} = -\epsilon m_{A'}^2 \vec{A'}$, which sources a magnetic field that can be calculated analytically for a rectangular volume with conductive boundary conditions~\cite{Chaudhuri:2014dla, Fedderke:2021aqo}. We assume the low-mass limit $m_{A'} \ll \hbar/(cL)$, so the dark photon wavelength is much longer than the room, allowing $\vec{A'} = (A'_x, A'_y, A'_z)$ to be treated as spatially constant.

We follow the standard treatment for a cavity driven by electric current,  see e.g. \cite{hill2009electromagnetic} and App.~A of \cite{Chaudhuri:2014dla}.  The conducting boundary condition implies $\hat n\times \vec E|_{\partial V}=0$ on the enclosure walls $\partial V$.
Working in the frequency domain with time dependence $e^{i\omega_{A'} t}$, the electric field can be expanded in a complete basis
\begin{equation}
\vec E(\vec r,t)=
\left(
\sum_{\alpha} c_{\alpha}\,\vec E_{\alpha}(\vec r)
+
\sum_{\beta} d_{\beta}\,\vec F_{\beta}(\vec r)
\right)e^{i\omega_{A'} t}.
\label{eq:E_decomp}
\end{equation}
The expansion consists of divergence-free eigenfunctions $\vec E_\alpha$ satisfying
\begin{equation}
\nabla^2 \vec E_{\alpha} = -\frac{\Omega_{\alpha}^2}{c^2}\,\vec E_{\alpha},\qquad
\nabla\cdot \vec E_{\alpha}=0,\qquad
\hat n\times \vec E_{\alpha}|_{\partial V}=0,
\end{equation}
and irrotational functions $\vec F_\beta=-\nabla \Phi_\beta$ with
\begin{equation}
\nabla^2 \Phi_{\beta} = -\frac{\tilde\Omega_{\beta}^2}{c^2}\,\Phi_{\beta},\qquad
\Phi_{\beta}|_{\partial V}=0,
\end{equation}
so that $\nabla\times \vec F_\beta=0$. We denote the cavity eigenvalues by $\Omega_\alpha$ to avoid confusion with the discrete time-domain Fourier frequencies $\omega_n$ used in Appendix~\ref{app:analysis}.

The coefficients $c_\alpha$ and $d_\beta$ follow by substituting Eq.~\eqref{eq:E_decomp} into the frequency-domain driven field equations, and then projecting onto each eigenfunction. Concretely, one multiplies by $\vec E_\alpha^{\,*}$ (or $\vec F_\beta^{\,*}$), integrates over the cavity volume, and uses orthogonality of the eigenfunctions  to obtain the standard driven-cavity expressions~\cite{hill2009electromagnetic}
\begin{align}
c_{\alpha} &=
\frac{-i\omega_{A'}}{\Omega_{\alpha}^2-\omega_{A'}^2}\,
\frac{\int_V d^3x\,\vec E_{\alpha}^{\,*}(\vec x)\cdot \vec J_{\rm eff}(\vec x)}
{\int_V d^3x\,|\vec E_{\alpha}(\vec x)|^2},
\label{eq:c_alpha}\\
d_{\beta} &=
\frac{i}{\omega_{A'}}\,
\frac{\int_V d^3x\,\vec F_{\beta}^{\,*}(\vec x)\cdot \vec J_{\rm eff}(\vec x)}
{\int_V d^3x\,|\vec F_{\beta}(\vec x)|^2}.
\label{eq:d_beta}
\end{align}
Using Faraday’s law $\nabla\times \vec E = -i\omega_{A'}\vec B$, the magnetic field is
\begin{equation}
\vec B(\vec r,t)=
\sum_{\alpha}
c_{\alpha}\,\frac{i}{\omega_{A'}}
\left(\nabla\times \vec E_{\alpha}(\vec r)\right)e^{i\omega_{A'} t},
\label{eq:B_from_modes}
\end{equation}
and receives no contribution from the $\vec F_\beta$ sector. We therefore only have to calculate the $c_\alpha$ coefficients to determine the magnetic field.
In our search band $\omega_{A'}\ll c/L$, the cavity is driven far off resonance, so $\omega_{A'}\ll \Omega_\alpha$ and
\begin{equation}
c_\alpha \simeq 
\frac{-i\omega_{A'}}{\Omega_\alpha^2}
\frac{\int_V d^3x\,\vec E_{\alpha}^{\,*}(\vec x)\cdot \vec J_{\rm eff}}
{\int_V d^3x\,|\vec E_{\alpha}(\vec x)|^2}
\end{equation}
up to corrections of order $(\omega_{A'}/\Omega_\alpha)^2$.

We now specialize to a rectangular volume $0<x<L_x$, $0<y<L_y$, $0<z<L_z$ with uniform
$\vec J_{\rm eff}=-\epsilon m_{A'}^2\vec A'$. In this geometry, the transverse electric eigenfunctions can be chosen as products of sines.
By linearity and cyclic permutation of $(x,y,z)$, it suffices to derive explicitly the field generated by a current along $\hat z$.
 Such a current excites modes with a single nonvanishing electric-field component $E_z(x,y)$ and no $z$ dependence,
\begin{equation}
E_{pq,z}(\vec x) \propto
\sin\!\Big(\frac{\pi p x}{L_x}\Big)
\sin\!\Big(\frac{\pi q y}{L_y}\Big).
\end{equation}
This is consistent with the conductive boundary condition because $E_z$ is normal to the $z=0,L_z$ walls (only the tangential components must vanish there). The corresponding eigenfrequencies are
\begin{equation}
\Omega_{pq}^{(z)\,2}=\pi^2 c^2\left(\frac{p^2}{L_x^2}+\frac{q^2}{L_y^2}\right),
\qquad p,q\in\mathbb{N}.
\label{eq:rect_eigenfreq_z}
\end{equation}
By cyclic permutation, analogous modes describe uniform drives along $\hat x$ and $\hat y$ with eigenfrequencies $\Omega_{pq}^{(x)}$ and $\Omega_{pq}^{(y)}$.
The overlap integral appearing in Eq.~\eqref{eq:c_alpha} is then
\begin{equation}
\int_V d^3x\,
\vec E_{pq}^{\,*}(\vec x)\cdot \vec J_{\rm eff}
=
J_z
\int_0^{L_x} dx
\int_0^{L_y} dy
\int_0^{L_z} dz\,
\sin\!\Big(\frac{\pi p x}{L_x}\Big)
\sin\!\Big(\frac{\pi q y}{L_y}\Big).
\end{equation}
Evaluating the integrals gives
\begin{equation}
\int_0^{L_x} dx\, \sin\!\Big(\frac{\pi p x}{L_x}\Big)
=
\begin{cases}
\displaystyle \frac{2L_x}{\pi p}, & p \text{ odd},\\
0, & p \text{ even},
\end{cases}
\end{equation}
and similarly for $q$. 

Plugging into Eq.~\eqref{eq:B_from_modes} and combining the contributions sourced by $A'_x$, $A'_y$, and $A'_z$ gives the magnetic field inside the rectangular enclosure~\cite{Chaudhuri:2014dla, Fedderke:2021aqo}:
\begin{equation} \begin{split} \label{eq:B_full}
	\vec{B} = -\frac{16\epsilon m_{A'}^2 c^2}{\pi^{3} \hbar^2} \sum_{p,q\text{ odd}}^{}\frac{1}{p q} 
	\Bigg( \frac{\frac{q}{L_y}\sin(\frac{\pi p x}{L_x})\cos(\frac{\pi q
	y}{L_y})}{ \frac{p^2}{L_x^2}+\frac{q^2}{L_y^2}}A'_z -
	\frac{\frac{p}{L_z}\cos(\frac{\pi p z}{L_z})\sin(\frac{\pi q
	x}{L_x})}{ \frac{p^2}{L_z^2}+\frac{q^2}{L_x^2}}A'_y,&\\
	 \frac{\frac{q}{L_z}\sin(\frac{\pi p y}{L_y})\cos(\frac{\pi q
	z}{L_z})}{ \frac{p^2}{L_y^2}+\frac{q^2}{L_z^2}}A'_x -
	\frac{\frac{p}{L_x}\cos(\frac{\pi p x}{L_x})\sin(\frac{\pi q
	y}{L_y})}{ \frac{p^2}{L_x^2}+\frac{q^2}{L_y^2}}A'_z,&\\
	\frac{\frac{q}{L_x}\sin(\frac{\pi p z}{L_z})\cos(\frac{\pi q
	x}{L_x})}{ \frac{p^2}{L_z^2}+\frac{q^2}{L_x^2}}A'_y -
	\frac{\frac{p}{L_y}\cos(\frac{\pi p y}{L_y})\sin(\frac{\pi q z}{L_z})}{
	\frac{p^2}{L_y^2}+\frac{q^2}{L_z^2}}A'_x&\bigg) ,  
\end{split} \end{equation}
where $L_{x,y,z}$ are the dimensions of the room and $(x, y, z)$ are the coordinates within it.

This formula justifies our choice of detector placement. Due to symmetry, each component of $\vec{A'}$ generates a magnetic field with a similar spatial pattern. Consider a current from $A'_x$: it generates magnetic fields only in the $y$ and $z$ directions, with an amplitude that is independent of the $x$ coordinate. Looking at the $yz$-plane, the field strength is maximal at the center of the walls. Thus, maximal sensitivity to $A'_x$ is achieved by placing the sensor anywhere along the center lines of the walls (e.g., at $y=L_y/2, z=0$ for any $x$). By choosing the specific location at the center of the floor ($x=L_x/2, y=L_y/2, z=0$), we simultaneously achieve maximal sensitivity to both the $A'_x$ and $A'_y$ components of the dark photon field.

The approximation provided in Eq.~\eqref{eq:B_scaling} of the main text was derived from the full expression by plugging the typical enclosure size $L_x\approx L_y=L_z\equiv L$, evaluating at the center of the floor $x=y=L/2,\, z=0$, and averaging over random directions of $\vec{A'}$. To make the linear scaling with $L$ explicit, note for example that the $A'_x$ contribution to $B_y$ at this point reduces to
\begin{equation}
B_y\Big(\frac{L}{2},\frac{L}{2},0\Big) = -\frac{16\epsilon m_{A'}^2 c^2}{\pi^{3} \hbar^2}\,A'_x\,L\sum_{p,q\text{ odd}}^{}\frac{(-1)^{\frac{p-1}{2}}}{p\left(p^2+q^2\right)} .
\label{eq:By_center_scaling}
\end{equation}

In the experiment the sensor is elevated by $\SI{20}{\centi\meter}$ and displaced by $\SI{15}{\centi\meter}$ along the $y$ axis from the exact geometric center of the floor. This breaks the exact symmetry and yields a small but nonzero dark-photon response in $B_z$. Evaluating Eq.~\eqref{eq:B_full} at this location gives a suppression of the expected $B_z$ signal by a factor of $\sim 20$ relative to the in-plane response, which is used in the noise-subtraction discussion in the main text when estimating possible signal leakage into the null channel.

\section{Analysis}
\label{app:analysis}

We describe here in detail the profile likelihood ratio test used to constrain the dark photon coupling parameter $\epsilon$. First, we define the likelihood function and explain how it is used to place a 95\% confidence level upper limit. Then, we define the threshold for identifying points where the white-noise-only hypothesis is rejected by the direct scan (before noise subtraction). Finally, we explain the veto tests used to check if these points are consistent with a dark photon signal.

\subsection{Likelihood and constraints}

\subsubsection{Definition}
The observed data, $\mathbf{d}$, consists of signal and noise contributions, $\mathbf{d} = \mathbf{S} + \mathbf{N}$. Here, $\mathbf{d}$ represents the Fourier components of the magnetic field measured in the $x$ and $y$ directions, $\mathbf{S}$ is the field generated by dark photons, and $\mathbf{N}$ is noise. To obtain the likelihood---the probability of observing the data $\mathbf{d}$ given the model parameters---we require the probability distributions of the signal, $P_S$, and the noise, $P_N$. The likelihood is then given by the expression
\begin{equation}
    L(\mathbf{d})=\int d \mathbf{S} P_{S} (\mathbf{S})
P_{\mathrm{N}}(\mathbf{d}- \mathbf{S}).
\end{equation}
The signal distribution, $P_s$, is derived from the statistical model of the dark photon field, while the noise distribution $P_N$ is assumed to be Gaussian and locally white.

Since the signal $\mathbf{S}$ is a function of the dark photon field $\bar{\mathbf{A}}$, we can change the variable of integration:
\begin{equation}
P_S(\mathbf{S})d\mathbf{S}=P_{\bar{A}}(\bar{\mathbf{A}}) d\bar{\mathbf{A}},
\end{equation}
where $P_{\bar{A}}$ is the probability distribution of $\mathbf{\bar{A}}$ given in Eq~\eqref{eq:A_distro}. The likelihood can therefore be written as
\begin{equation} \label{eq:L_integral}
    L(\mathbf{d})=\int d\bar{\mathbf{A}} P_{\bar{A}}(\bar{\mathbf{A}})
P_{\mathrm{N}}(\mathbf{d}-\alpha \bar{\mathbf{A}}),
\end{equation}
where $\mathbf{S} = \alpha\bar{\mathbf{A}}$. The matrix $\alpha$ describes the linear transformation from the dark photon field to the measured magnetic fields, with its elements determined by the magnetic calibration measurements, as well as Eqs.~\eqref{eq:B_full} and \eqref{eq:A_normalization}. We now specify the noise distribution $P_N$ to obtain the final likelihood expression.

\subsubsection{Noise model}
We assume the noise to be Gaussian and locally white. The noise variance is constant within the signal bandwidth, but allowed to be  different for the $x$ and $y$ axes. The distribution is given by:
\begin{equation} \label{eq:N_distro}
    P_N(\mathbf{N}) = \frac{1}{(2\pi)^{n_N}|\Sigma_N|}
    e^{-\mathbf{N}^{\dagger}\Sigma_N^{-1}\mathbf{N}} = 
     \frac{1}{(2\pi)^{n_f}\sigma_x^{2n_f}}
    e^{-\mathbf{N}_x^{\dagger}\mathbf{N}_x/\sigma_x^2}
     \frac{1}{(2\pi)^{n_f}\sigma_y^{2n_f}}
    e^{-\mathbf{N}_y^{\dagger}\mathbf{N}_y/\sigma_y^2}.
\end{equation}
Here $\mathbf{N}_{x,y}$ are the noise components and $\sigma_{x,y}^2$ are the corresponding variances. The noise covariance matrix, $\Sigma_N$, is diagonal, with blocks corresponding to the $x$ and $y$ directions, where each block is the identity matrix multiplied by $\sigma_x^2$ or $\sigma_y^2$. The analysis includes $n_f$ Fourier frequencies, so the total number of noise components is $n_N = 2n_f$.

The noise variances, $\sigma_{x,y}^2$, are estimated empirically from the data in frequency sidebands adjacent to the signal region. For each frequency analyzed, we define sidebands on either side, each containing 10 times the number of frequency bins as the signal region itself. Within these sidebands, we compute the median of the squared norm of the Fourier components. The variance is then estimated by multiplying this median by 2.2, which is the theoretical ratio of variance-to-median for the power of a Gaussian-distributed variable. We use the median rather than the mean to make our noise estimate robust against narrow spectral artifacts. In the absence of such artifacts, this method yields a result very similar to a simple average.

\subsubsection{Final likelihood}
Plugging the distributions from Eqs.~\eqref{eq:A_distro} and \eqref{eq:N_distro} into the integral in Eq.~\eqref{eq:L_integral}, we get
\begin{equation}
    L(\mathbf{d})=\int d\bar{\mathbf{A}} 
    \frac{1}{(2\pi)^{4n_f}|\Sigma_N|
    \left|\Sigma_{\bar{A}}\right|} e^{-\bar{\mathbf{A}}^{\dagger}
\Sigma_{\bar{A}}^{-1} \bar{\mathbf{A}} / 2 -\left( \mathbf{d}-\alpha \bar{\mathbf{A}}\right)
^{\dagger}\Sigma_N^{-1}\left( \mathbf{d}-\alpha \bar{\mathbf{A}}\right) /2},
\end{equation}
where we have used $n_{\bar{A}} = n_N = 2n_f$. We can solve this using the generic Gaussian integration formula
\begin{equation}
\int d^n x e^{-\mathbf{x}^{\dagger} M \mathbf{x} / 2+\mathbf{J}^{\dagger}
\mathbf{x} / 2+\mathbf{x}^{\dagger} \mathbf{J} / 2}=\frac{(2 \pi)^n}{|M|}
e^{\mathbf{J}^{\dagger} M^{-1} \mathbf{J} / 2}
\end{equation}
with the identifications
\begin{equation}
\begin{gathered} 
    M = \Sigma_{\bar{A}}^{-1} + \alpha ^{\dagger}\Sigma_N ^{-1}\alpha, \\
    \mathbf{J} = \alpha^{\dagger}\Sigma_N^{-1}\mathbf{d}
\end{gathered}
\end{equation}
to obtain the final likelihood expression:
\begin{equation}
    L(\mathbf{d})=
    \frac{1}{(2\pi)^{2n_f}|\Sigma_N| \left|\Sigma_{\bar{A}}\right|
    \left| M \right
    |} 
    e^{-\mathbf{J}^{\dagger} M^{-1} \mathbf{J} / 2
    -\mathbf{d}^{\dagger} \Sigma_N^{-1} \mathbf{d} /2} .
\end{equation}

\subsubsection{Placing constraints}
With the likelihood defined, we use the profile likelihood ratio test to find the 95\% confidence level upper limit on the mixing parameter $\epsilon$ as a function of mass. For each mass $m_{A'}$, we maximize the likelihood with respect to $\epsilon$, finding its maximum likelihood estimate, $\hat{\epsilon}$. From this value, $\epsilon$ is increased until it reaches the point where
\begin{equation}
    -2\log \left(\frac{L(\epsilon,m_{A'})}{L(\hat{\epsilon},m_{A'})}\right) = 2.71.
\end{equation}
This value of $\epsilon$ is taken as the upper limit. The critical value of 2.71 is based on the asymptotic half-$\chi^2$ distribution~\cite{Cowan:2010js}.

\subsection{Coherent peak detection}
\subsubsection{Threshold definition}
The criteria for rejecting the white-noise-only hypothesis is based on the requirement of equivalence to a $5\sigma$ Gaussian significance, leading to a $p$-value of $5.7\times10^{-7}$. However, our analysis was performed for millions of independent points, making for a significant look-elsewhere effect~\cite{Gross:2010qma}. To mitigate this look-elsewhere effect, we lowered the $p$-value threshold by a factor of $10^{-7}$. Plugging the reduced $p$-value into the half-$\chi^2$ distribution, we obtained the threshold
\begin{equation} \label{eq:detection_threshold}
        -2\log \left(\frac{L(0,m_{A'})}{L(\hat{\epsilon},m_{A'})}\right) \geq 56,
\end{equation}
with $\hat{\epsilon}$ being the maximum likelihood estimate of $\epsilon$ as above. Masses satisfying this inequality in the direct (un-subtracted) scan were identified as the points where the white-noise hypothesis is rejected.

\subsubsection{Vetoing tests}
Points identified as including a signal, as mentioned above, were passed through three tests designed to be passed by signals created by a true dark photon.
We now explain with technical detail how the tests were performed. 
\begin{enumerate}
    \item \textbf{Spectral Shape:} Similar to previous axion searches~\cite{Bloch:2022kjm}, this test evaluates the candidate's spectral shape. To quantitatively compare the spectral shape to a true dark photon signal, we have simulated this true signal and evaluated it's predicted peak height compared to the observed peak.
    
    Specifically, a dark photon signal was simulated a thousand times, with amplitude corresponding to $\hat{\epsilon}$, the maximum likelihood value. The ensemble of simulations provided a distribution for the peak amplitude of the signal in the $y$ direction (an arbitrary choice, the $x$ signal could be used as instead). From the distribution, the 95\% quantile was extracted, and compared to the observed peak. By definition, a true dark photon signal would have a 95\% probability of passing this test--but a large fraction of the observed peaks didn't, indicating they are not sourced by dark photons. 

    \item \textbf{Coincident Signal in $\boldsymbol{B_z}$:} In this test, the $B_z$ channel is searched for a signal inconsistent with a dark photon origin. In practice, the position of the device is not precisely the center of the floor, so a small signal is predicted to be sourced by a true dark photon. However, even with the imprecise positioning, the predicted amplitude from a true dark photon is smaller than the $x$ and $y$ amplitudes by more than an order of magnitude. Namely, with the exact positioning in our experiment, the $z$-axis magnetic field calculated with Eq.~\eqref{eq:B_full} is smaller than the $x$-axis field by a factor of 20. Therefore, an amplitude in the $z$ direction comparable to the other directions indicates it is not originating from a dark photon.
    
    Quantitatively, the test was performed by running an independent maximum likelihood fit to the $z$ axis data, searching it for a dark photon signal.
    The search on the $z$-axis data yields a corresponding maximum likelihood value of the mixing parameter, $\hat{\epsilon}_z$. The presence of a stronger than predicted $B_z$ signal is indicated by $\hat{\epsilon}_z>\hat{\epsilon}$, i.e. the value of the mixing parameter obtained from the $B_z$ is above the one obtained from $B_x$ and $B_y$. Points satisfying this condition were rejected if the value of $\hat{\epsilon}$ was outside the $95\%$ confidence level measurement of $B_z$:
    \begin{equation}
        -2\log \left(\frac{L_z(5\,\hat{\epsilon},m_{A'})}{L_z(\hat{\epsilon_z},m_{A'})}\right) > 2.71,
    \end{equation}
    where we have added a factor of 5 for robustness of the test against inaccuracies in the ratio of calculation of $B_z$, which is expected to be lower than $B_x$ by a factor of 20 as mentioned above. 
    An inaccuracy in this factor is expected because the signal strength is sensitive to small changes in the device position, which is close the the point where the signal vanishes (see Eq.~\eqref{eq:B_full}).

    \item \textbf{Post-Subtraction Significance:} This test uses the enhanced sensitivity from our noise-subtraction analysis. If a candidate survives the first two vetoes, we check it against the more stringent 95\% confidence limit derived from the cleaned data. If the candidate's $\epsilon$ value is above this improved limit, it is rejected.
\end{enumerate}

\subsubsection{Remaining point}
In the initial scan, a bit more than 70,000 mass values were identified as satisfying the criterion defined in Eq.~\eqref{eq:detection_threshold}. Of all these candidates, only a single one---corresponding to the mass $m_{A'}c^2/h = 1090.0303\,\mathrm{Hz}$---passed all three veto tests.

\begin{figure}[htpb]
	\centering
	\includegraphics[width=0.8\textwidth]{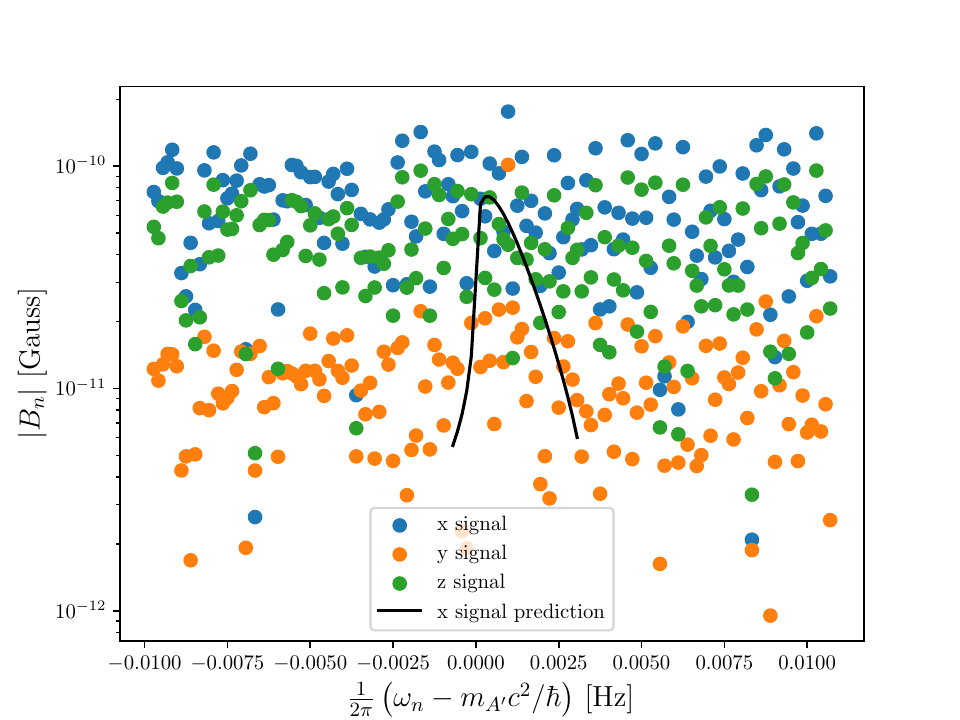}
	\caption{
    Magnitude of the Fourier-transformed magnetic field in the vicinity of the surviving point at $m_{A'}c^2/h = 1090.0303\,\mathrm{Hz}$. 
The scattered points show the absolute values of the Fourier coefficients in the $x$, $y$, and $z$ directions. 
The solid black line indicates the predicted signal amplitude in the $x$ direction for a dark photon with mixing parameter equal to the candidate’s maximum-likelihood value; the predicted $y$ amplitude has an identical shape but is lower by approximately $33\%$ and is not shown for visual clarity. 
Only a single frequency component in each of the $x$ and $y$ spectra lies noticeably above the surrounding noise.
    }
	\label{fig:detection}
\end{figure}

Fig.~\ref{fig:detection} shows the data in the vicinity of this mass. The scattered points display the measured absolute values of the Fourier-transformed magnetic field components, while the black line indicates the expected signal amplitude in the $x$ direction for a dark photon with mixing parameter equal to the candidate’s maximum-likelihood value. The predicted $y$-component has a similar profile but an amplitude lower by approximately $33\%$.

A visual inspection shows that both the $x$ and $y$ spectra contain a single frequency bin whose amplitude exceeds the surrounding noise. In an idealized scenario, such a feature would typically be rejected by the spectral-shape veto, which tests consistency with the broader and smoother lineshape expected from a dark photon signal. The survival of this point is attributed to its low mass, for which the expected signal bandwidth spans only a small number of discrete Fourier bins, rendering the spectral-shape test less discriminating. This point requires further examination beyond our pre-defined vetoes to test the nature of its origin.
\FloatBarrier

\section{Shielded Room Magnetic Noise Attenuation}
\label{app:Room}
The magnetic shielding effectiveness of the room was characterized over the 80--180~kHz frequency range by measuring the attenuation of a controlled magnetic field.
The excitation field was generated by a copper induction coil with 21 turns of 15 cm diameter, driven by a Keysight 33200A waveform generator at constant voltage mode. The resulting drive current, inferred from the coil impedance, assumed unchanged between configurations. The magnetic field was detected using the sensor described in the main text.
A reference measurement was first acquired with both the coil and the sensor positioned inside the room at a fixed separation of 2 m and fixed relative orientation. The measurement was then repeated with the coil placed outside the room while preserving the same geometry with respect to the sensor. The excitation frequency was swept across 80--180~kHz using single-tone steps. 

 \begin{figure}[htpb]
    \centering \includegraphics[width=0.7\linewidth]{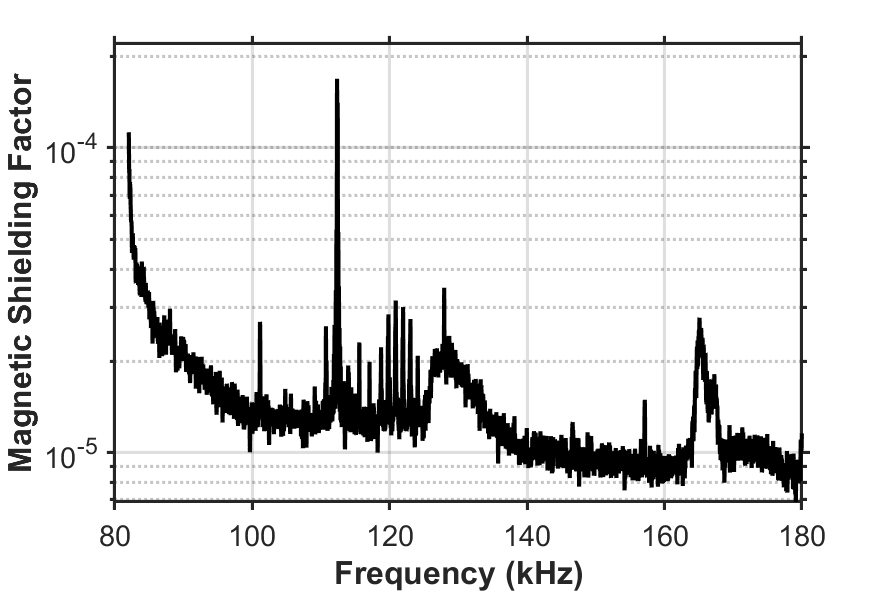} \caption{Measured conservative bound on the magnetic shielding factor in the frequency range 80--180~kHz, expressed as the ratio of magnetic field amplitude measured with the source outside the room to that measured with the source inside. Deviations below $\sim$100~kHz and narrow spectral features are caused by the limitation of the measured noise floor in the room.} 
 \label{fig:room_attenuation} 
 \end{figure}
 
Across the scanned spectrum, the magnetic field amplitude measured inside the room remained at or below the measured noise floor. This establishes a conservative bound on the attenuation of magnetic field Amplitude Spectral Density (ASD) of order $10^{-5}$. Figure~\ref{fig:room_attenuation} shows the ratio of magnetic field amplitude spectral densities (ASDs) measured with the source outside and inside the room, representing the measured minimum attenuation.
To verify that the measurement was limited by the instrumental noise floor, the coil–sensor separation was reduced to 50~cm. The measured signal remained noise-floor limited despite the expected fourfold increase in field amplitude. 
The apparent rise in the ratio below $\sim$100~kHz and the narrow spectral features are attributed to limitations imposed by the electronic noise floor rather than physical magnetic leakage. The actual shielding performance is therefore expected to exceed the reported bound; however, further quantification was limited by the maximum available drive amplitude.

\newpage
\bibliographystyle{ieeetr}
\bibliography{NASDUCK_references}

\end{document}